\documentclass[12pt]{elsarticle}

\usepackage{amsmath, amsthm, amssymb, amsfonts}
\usepackage{graphicx}
\usepackage{float}
\usepackage[hidelinks]{hyperref}
\usepackage[utf8]{inputenc}
\usepackage[english]{babel}
\usepackage{svg}
\usepackage{caption}
\usepackage{subcaption}
\usepackage{multirow}

\journal{arXiv}

\begin{document}

\newtheorem{df}{Definition}
\newtheorem{prop}{Proposition}
\newtheorem{thm}{Theorem}
\newtheorem{cor}{Corollary}
\newtheorem{as}{Assumption}
\newtheorem{fa}{Fact}

\begin{frontmatter}



\title{Modelling and valuation of catastrophe bonds across multiple regions}






\author[inst1]{Krzysztof Burnecki}
\author[inst1]{Marek Teuerle}
\author[inst1]{Martyna Zdeb}

\affiliation[inst1]{organization={Faculty of Pure and Applied Mathematics, Hugo Steinhaus Center, Wroc{\l}aw University of Science and Technology},
            addressline={Wyspianskiego~27},
            city={50-370 Wroc{\l}aw},
            country={Poland}}

\begin{abstract}
The insurance-linked securities (ILS) market, as a form of alternative risk transfer, has been at the forefront of innovative risk-transfer solutions.  The catastrophe bond (CAT bond) market now represents almost
half of the entire ILS market and is growing steadily. Since CAT bonds are often tied to risks in different regions, we follow this idea by constructing different pricing models that incorporate various scenarios of dependence between catastrophe losses in different areas.  Namely, we consider independent, proportional, and arbitrary two-dimensional distribution cases. We also derive a normal approximation of the prices. Finally, to include the market price of risk, we apply Wang's transform. We illustrate the differences between the scenarios and the performance of the approximation on the Property Claim Services data.   

\end{abstract}




\end{frontmatter}

\section{Introduction}
Floods, earthquakes, and severe storms are major threats to society. In addition to loss of life and bodily injury, natural catastrophes can inflict damage to property, reducing both wealth and productive capacity, potentially resulting in significant financial losses on both macro- and micro-levels. The total cost typically reflects both the severity of the initial damage and the speed with which reconstruction can be completed. This is why insurance can play a protective role \cite{Swissres}.

Insured losses from global natural catastrophes reached USD 137 billion in 2024, driven primarily by Hurricanes Helene and Milton and severe convective storms. Secondary perils were the largest contributor to these losses, and significant fires, such as those in Los Angeles in early 2025, suggest that this trend of high secondary peril losses may continue. Looking ahead, if the typical 5-7\% annual increase persists, insured losses could reach USD 145 billion in 2025. Although secondary risks have recently represented a larger part of yearly insured losses, primary risks still represent the greatest potential for catastrophic loss. Historically, a single major primary event, such as a hurricane hitting a densely populated urban centre, could result in losses that far exceed the current trend \cite{Swissreport2025}.

Developed in the 1990s, catastrophe bonds (CAT bonds) are a financial tool that allows insurance companies to offload natural disaster risks onto capital markets. They were created as an innovative answer to the growing number and intensity of major events such as hurricanes, earthquakes, and floods, which had significantly stressed insurers and reinsurers financially. For investors, these bonds offer a way to diversify their holdings while simultaneously providing insurers with protection against massive, potentially devastating losses \cite{Barrieu2010,Barrieu2025}.

The structure commonly involves a Special Purpose Vehicle (SPV), which issues the bonds to investors and simultaneously enters into a reinsurance agreement with the sponsor. The premium paid by the sponsor to the SPV funds the coupon payments made to the bondholders. Payouts to the sponsor (and corresponding losses of principal for investors) are contingent upon the occurrence of a predefined catastrophic event exceeding a certain severity threshold (usually three years), determined by specific trigger mechanisms. These triggers can be based on the sponsor's actual losses (indemnity), objective physical parameters of the event (parametric), industry-wide losses (industry loss), or losses estimated by a third-party model (modelled loss) \cite{BraunKousky2021}.

CAT bonds function as a type of Alternative Risk Transfer, offering advantages not just to insurers but also to governments and businesses highly exposed to natural disasters. Occasionally, national entities have issued these bonds to secure funds for post-disaster relief and recovery. As such, CAT bonds are a crucial instrument for enhancing resilience in areas susceptible to natural catastrophes, helping to stabilise local economies after such events. In addition, pioneering risk transfer methods, such as CAT bonds, can help address existing protection gaps and improve individual and community resilience \cite{burnecki2023catastrophe}.

For the second year in a row, more than 90 catastrophe bond and related insurance-linked transactions were placed in 2024, with the 93 coming close to the 95 deals of last year. In terms of 144A property cat bond issuance, 2024 has set a new record of USD 16.6 billion, up 11\% on the previous high of 2023. The total 144A cat bond issuance, which includes 144A cat bonds Covering other lines of re/insurance business such as cyber and terrorism, also hit a new high of USD 17.2 billion \cite{artemisreport2025}. 



Much of the research focused on modelling natural catastrophes and pricing related risk securitization solutions addresses the issue of arbitrage-free valuation and market completeness. The field evolved from using simple early models to more complex approaches incorporating stochastic interest rates and contingent claims \cite{taylor,burkuktay11,Brenda2010,mm,NowakRomaniuk2013,NowakRomaniuk2018,vaugirard2004canonical,braun}. Currently, there is increased interest in pricing advanced instruments like catastrophe options, futures, contingent convertible CAT bonds, and risk swaps  \cite{braun,burgiupal2019,Arnone2021Catastrophic}.

We also note that to address the pricing challenges in insurance and finance, particularly for non-hedgeable risks, various risk measures and premium calculation principles have been developed. One important class is distortion risk measures  explored by Wang \cite{Wang1995, Wang1996a}. These measures calculate an insurance premium or risk capital by transforming, or distorting, the probability distribution of the underlying loss variable \cite{Wang2002}. 

In practice, the issuance of a catastrophe bond typically involves a specialized modeling firm to quantify the catastrophe risk. These risk modelers estimate the probability of first loss (PFL) and provide an expected loss (EL) estimate for investors. Several studies have examined the relationship between price spreads and EL and other explanatory variables in the regression framework, see, e.g. \cite{braun2,LaneBeckwith2021}. Additional work includes studies on implied Poisson intensities from observed prices \cite{BEER2022106350}, the use of machine learning techniques \cite{Lane2018,GotGurWit2020}, the application of mean utility functionals for CAT bond pricing \cite{petra2021}, and the use of extreme value theory for pricing earthquake-related catastrophe bonds \cite{zimbidis2007modeling}.

Historically, CAT bonds were designed with a single trigger event that, if it occurs and meets specified criteria, can lead to a payout to the issuer and potential loss of principal for investors. However, the use of multiple triggers is gaining importance. CAT bonds are frequently structured into several tranches. These tranches are differentiated by their risk-return profiles. The variation between tranches can be based on several factors, including the specific trigger event, the reference peril (the type of natural catastrophe Covered), and the defined Covered territory or geographic area. This allows different tranches to cater to investors with varying risk appetites. Existing literature has  highlighted the issue of multi-peril ILS sometimes related to different regions. See, for instance, the pioneering work by Lane \cite{lane2004} explored "arbitrage-equivalent" pricing for Covers that could be either bought or sold or \cite{haslip2010pricing}, a study divided Italy into three seismic zones and priced three CAT bonds with different default risk levels, and also \cite{Shao2015} for an empirical study of
California earthquake data within the multi-peril risk model.  In \cite{Burnecki2024Pricing} the authors introduced a model for multi-peril CAT bonds where a trigger can be activated if losses from at least one Covered peril exceed a set threshold.

Reflecting the multi-regional nature of CAT bond risks, in this paper, we introduce pricing models which are designed to incorporate diverse scenarios regarding the dependence of catastrophe losses across different geographic areas.


\section{Three approaches for modelling dependence in different regions}
In this section we introduce three approaches to modelling natural catastrophe losses in two regions. The methodology can be extended to more than two regions. The first assumes full independence between the aggregate loss processes in both regions. In the second case, we assume that the loss amounts of common catastrophes split between between the two regions with a fixed (deterministic) proportion. Other catastrophes that hit separately the two regions are modelled with independent aggregate loss processes. The third approach assumes that the proportion for common catastrophes varies randomly from loss to loss.

\subsection{Model 1. Independent loss processes}
We want to model losses caused by a selected peril in two regions. The simplest idea is to model them as two independent aggregate loss processes $(S_i(t), t>0), i=1,2$, each corresponding to one of the regions:
\everymath={\displaystyle}
\begin{align}
\left\{
	\begin{array}{ll}
		 S_1(t) &= \sum\limits_{i=1}^{N^{(1)}(t)} X_i,\\ \\
	   S_2(t) &= \sum\limits_{i=1}^{N^{(2)}(t)} Y_i, 
	\end{array}
\right.
\label{eq:independent losses}
\end{align}
where $N^{(1)}(t)$ and $N^{(2)}(t)$ are counting processes, independent of each other, and $\{X_i, i \in \mathbb{N}\}$ and $\{Y_i, i \in \mathbb{N}\}$ are independent identically distributed (i.i.d.) positive random variables describing the losses in each region.
We also assume that $\{X_i, i \in \mathbb{N}\}$ and $\{Y_i, i \in \mathbb{N}\}$ are independent.

\subsection{Model 2. Losses split proportionally with a fixed proportion}
For the second approach, we assume that when a catastrophe hits both of the considered regions, the losses are split with constant proportion~$p$, $0< p <1$. The first region is affected by the part $p$ of the loss and the part $1-p$ affects the second region. The model can be written as:
 \everymath={\displaystyle}
\begin{align}
\left\{
	\begin{array}{ll}
		 S_1(t) &= \sum\limits_{i=1}^{N^{(1)}(t)} X_i + p\sum\limits_{i=1}^{N^{(2)}(t)} Z_i, \\ \\
	   S_2(t) &= \sum\limits_{i=1}^{N^{(3)}(t)} Y_i + (1-p)\sum\limits_{i=1}^{N^{(2)}(t)} Z_i. 
	\end{array}
\right.
\label{eq:proportional losses}
\end{align}

The processes $N^{(1)}(t)$ and $N^{(3)}(t)$ count losses that occurred only in one of the regions, and the i.i.d. variables $\{X_i, i\in \mathbb{N}\}$ and $\{Y_i, i\in \mathbb{N}\}$ describe the losses in the first and second regions, respectively. The process $N^{(2)}(t)$ describes the flow of losses that appear in both regions.  The variables $\{Z_i, i\in \mathbb{N}\}$ represent the corresponding losses and are divided in proportion $p$ between two regions. We assume that all counting processes and all loss amount random variables are mutually independent.

\subsection{Model 3. Dependent losses}
Instead of splitting common losses with a fixed proportion, we can model them as correlated random variables with a given correlation coefficient. In that case, the two-dimensional aggregate loss process that describes the losses in the considered regions can be written as:
\everymath={\displaystyle}
\begin{align}
\left\{
	\begin{array}{ll}
		 S_1(t) &= \sum\limits_{i=1}^{N^{(1)}(t)} X_i^{(1)} + \sum\limits_{i=1}^{N^{(2)}(t)} X_i^{(2)}, \\ \\
	   S_2(t) &= \sum\limits_{i=1}^{N^{(3)}(t)} Y_i^{(1)} + \sum\limits_{i=1}^{N^{(2)}(t)} Y_i^{(2)}. 
	\end{array}
\right.
\label{eq:dependent losses}
\end{align}

The part describing losses only in one of the regions is the same as in Model 2, 
namely the processes $N^{(1)}(t)$ and $N^{(3)}(t)$ count losses that occurred only in one of the regions, and i.i.d. variables $\{X_i^{(1)}, i\in \mathbb{N}\}$ and $\{Y_i^{(1)}, i\in \mathbb{N}\}$ model the sizes of these losses. The process $N^{(2)}(t)$ counts losses that occur in both regions at the same time and variables $\{X_i^{(2)}, i\in \mathbb{N}\}$ and $\{Y_i^{(2)}, i\in \mathbb{N}\}$  model their sizes. We assume that they are, in general, dependent and we know their correlation coefficient. 

We assume that all counting processes $N^{(i)},i=1,2,3$ are independent of loss amount random variables and that the losses that occured only in one region are independent of the losses from catastrophes that hit both regions.

\section{Pricing of a zero-coupon CAT bond under a physical measure}
Let $\mathbf{S} = \{S_1(t), S_2(t)\}$ be a two-dimensional process describing losses caused by a chosen peril in two regions. 
A zero-coupon (ZC) CAT bond with the maturity time $T$ will be triggered if the aggregate losses of at least one of the regions exceed corresponding threshold:
\everymath={\displaystyle}
\begin{align}
P(T) = 
\left\{
	\begin{array}{ll}
		 1, & \mathrm{if} \left(S_1(T) < D_1 \land S_2(T) < D_2\right), \\
	   c, & \mathrm{if} \left(S_1(T) \geq D_1 \lor S_2(T) \geq D_2\right), 
	\end{array}
\right.
\label{eq:ZC CAT payoff}
\end{align}
where $0\leq c \leq 1$ is a recovery rate and $D_i$ ($i=1,2$) denotes the threshold for the respective region.

The price of a ZC CAT bond with maturity time $T$ and face value $1$, with payoff function defined in equation (\ref{eq:ZC CAT payoff}) can be written as:
\begin{align}
    V_0&=e^{-rT}\mathrm{E}_{\mathbb{P}}\left[P(T)\right]=\nonumber \\[5pt] 
    &=e^{-rT}\left[\mathrm{P}\left(S_1(T) < D_1 \land S_2(T) < D_2\right)\right. \\
    & \qquad + c \left.\mathrm{P}\left(S_1(T) \geq D_1 \lor S_2(T) \geq D_2\right) \right]=  \nonumber \\[5pt]
    &=e^{-rT} \left[c+(1-c)\mathrm{P}\left(S_1(T) < D_1 \land S_2(T) < D_2\right)\right], \label{eq:ZC CAT price}
\end{align}
where $\mathbb{P}$ is the physical measure.

\subsection{Normal approximation}
To balance tractability, speed, and accuracy, actuaries can apply approximations to solve real-world insurance problems. We propose here a normal approximation of the price of the considered bond.

We want to approximate the price given by formula \eqref{eq:ZC CAT price} using the bivariate normal distribution $\boldsymbol{N}=(N_1,N_2)$ with mean 
\begin{equation}
\boldsymbol{\mu} = 
    \begin{pmatrix}
        \mathrm{E}S_1\left(T\right)\\
        \mathrm{E}S_2\left(T\right)
    \end{pmatrix}
\end{equation}
and covariance matrix:
\begin{equation}
\boldsymbol{\Sigma}= 
    \begin{pmatrix}
        \mathrm{Var}S_1\left(T\right) & \mathrm{Cov}\left(S_1\left(T\right), S_2\left(T\right)\right)\\
        \mathrm{Cov}\left(S_1\left(T\right), S_2\left(T\right)\right) & \mathrm{Var}S_2\left(T\right)
    \end{pmatrix}.
\end{equation}
This leads to the approximate pricing formula:
\begin{equation}
    V_0^{approx} = e^{-rT} \left[c+(1-c)\mathrm{P}\left(N_1 < D_1 \land N_2 < D_2\right)\right]. \label{eq:ZC CAT normal approximation}
\end{equation}
We now aim to specify the formula for Models 1-3.


\subsubsection{Model 1}
For Model 1 given by equation (\ref{eq:independent losses}) 
\begin{equation}
\boldsymbol{\mu} = 
    \begin{pmatrix}
        \mathrm{E}S_1\left(T\right)\\
        \mathrm{E}S_2\left(T\right)
    \end{pmatrix}
    =
    \begin{pmatrix}
        \mathrm{E}\left(N^{(1)}(T)\right)\mathrm{E}X_1 \\
        \mathrm{E}\left(N^{(2)}(T)\right)\mathrm{E}Y_1 
    \end{pmatrix}\label{eq:na ind mu}
\end{equation}
and covariance matrix
\begin{equation}
\boldsymbol{\Sigma}= 
    \begin{pmatrix}
    \mathrm{Var}S_1\left(T\right) & \mathrm{Cov}\left(S_1\left(T\right), S_2\left(T\right)\right)\\
        \mathrm{Cov}\left(S_1\left(T\right), S_2\left(T\right)\right) & \mathrm{Var}S_2\left(T\right)
    \end{pmatrix},\label{eq:na ind sig}
\end{equation}
where
\begin{align*}
    \mathrm{Var}&\left(S_1\left(T\right)\right)=\mathrm{Var}(X_1) \mathrm{E}\left(N^{(1)}(T)\right) + \left(\mathrm{E}(X_1)\right)^2 \mathrm{Var}\left(N^{(1)}(T)\right),\nonumber\\[10pt]
    \mathrm{Var}&\left(S_2\left(T\right)\right)=\mathrm{Var}(Y_1) \mathrm{E}\left(N^{(2)}(T)\right) + \left(\mathrm{E}(Y_1)\right)^2 \mathrm{Var}\left(N^{(2)}(T)\right),\nonumber\\[10pt]
    \mathrm{Cov}&\left(S_1\left(T\right), S_2\left(T\right)\right) = 0.
\end{align*}

\subsubsection{Model 2}
For Model 2 given by equation (\ref{eq:proportional losses})
\begin{equation}
\boldsymbol{\mu} = 
    \begin{pmatrix}
        \mathrm{E}S_1\left(T\right)\\
        \mathrm{E}S_2\left(T\right)
    \end{pmatrix}
    =
    \begin{pmatrix}
        \mathrm{E}\left(N^{(1)}(T)\right)\mathrm{E}X_1 + p\,\mathrm{E}\left(N^{(2)}(T)\right)\mathrm{E}Z\\
        \mathrm{E}\left(N^{(3)}(T)\right)\mathrm{E}Y_1 + (1-p)\,\mathrm{E}\left(N^{(2)}(T)\right)\mathrm{E}Z
    \end{pmatrix},
\end{equation}
and covariance matrix
\begin{equation}
\boldsymbol{\Sigma}= 
    \begin{pmatrix}
        \mathrm{Var}S_1\left(T\right) & \mathrm{Cov}\left(S_1\left(T\right), S_2\left(T\right)\right)\\
        \mathrm{Cov}\left(S_1\left(T\right), S_2\left(T\right)\right) & \mathrm{Var}S_2\left(T\right)
    \end{pmatrix},
\end{equation}
where
\begin{align*}
    \mathrm{Var}&\left(S_1\left(T\right)\right)=\mathrm{Var}(X_1) \mathrm{E}\left(N^{(1)}(T)\right) + \left(\mathrm{E}(X_1)\right)^2 \mathrm{Var}\left(N^{(1)}(T)\right)+\nonumber\\[5pt]
    & +p^2 \left[\mathrm{Var}(Z_1) \mathrm{E}\left(N^{(2)}(T)\right)+\left(\mathrm{E}Z_1\right)^2\mathrm{Var}\left(N^{(2)}(T)\right)\right], \\[10pt]
    \mathrm{Var}&\left(S_2\left(T\right)\right)=\mathrm{Var}(Y_1) \mathrm{E}\left(N^{(3)}(T)\right) + \left(\mathrm{E}(Y_1)\right)^2 \mathrm{Var}\left(N^{(3)}(T)\right)+\nonumber\\[5pt]
    & +(1-p)^2 \left[\mathrm{Var}(Z_1) \mathrm{E}\left(N^{(2)}(T)\right)+\left(\mathrm{E}Z_1\right)^2\mathrm{Var}\left(N^{(2)}(T)\right)\right], \\[10pt]
    \mathrm{Cov}&\left(S_1\left(T\right), S_2\left(T\right)\right) = p(1-p)\left[ \mathrm{Var}(Z_1) \mathrm{E}\left(N^{(2)}(T)\right)\right.\\
    &+ \left.\left(\mathrm{E}Z_1\right)^2 \mathrm{Var}\left(N^{(2)}(T)\right) \right].
\end{align*}

\subsubsection{Dependent losses}
For Model 3 given by equation (\ref{eq:dependent losses})
\begin{equation}
\boldsymbol{\mu} = 
    \begin{pmatrix}
        \mathrm{E}S_1\left(T\right)\\
        \mathrm{E}S_2\left(T\right)
    \end{pmatrix}
    =
    \begin{pmatrix}
        \mathrm{E}\left(N^{(1)}(T)\right)\mathrm{E}X^{(1)} + \mathrm{E}\left(N^{(2)}(T)\right)\mathrm{E}X^{(2)}\\
        \mathrm{E}\left(N^{(3)}(T)\right)\mathrm{E}Y^{(1)} + \mathrm{E}\left(N^{(2)}(T)\right)\mathrm{E}Y^{(2)}
    \end{pmatrix}
\end{equation}
and covariance matrix
\begin{equation}
\boldsymbol{\Sigma}= 
    \begin{pmatrix}
        \mathrm{Var}S_1\left(T\right) & \mathrm{Cov}\left(S_1\left(T\right), S_2\left(T\right)\right)\\
        \mathrm{Cov}\left(S_1\left(T\right), S_2\left(T\right)\right) & \mathrm{Var}S_2\left(T\right)
    \end{pmatrix},
\end{equation}
where
\begin{align*}
    \mathrm{Var}&\left(S_1\left(T\right)\right)=\mathrm{Var}(X^{(1)}) \mathrm{E}\left(N^{(1)}(T)\right) + \left(\mathrm{E}(X^{(1)})\right)^2 \mathrm{Var}\left(N^{(1)}(T)\right)+\nonumber\\[5pt]
    &\qquad+\mathrm{Var}(X^{(2)}) \mathrm{E}\left(N^{(2)}(T)\right) + \left(\mathrm{E}(X^{(2)})\right)^2 \mathrm{Var}\left(N^{(2)}(T)\right), \\[10pt]
    \mathrm{Var}&\left(S_2\left(T\right)\right)=\mathrm{Var}(Y^{(1)}) \mathrm{E}\left(N^{(3)}(T)\right) + \left(\mathrm{E}(Y^{(1)})\right)^2 \mathrm{Var}\left(N^{(3)}(T)\right)+\nonumber\\[5pt]
    &\qquad +\mathrm{Var}(Y^{(2)}) \mathrm{E}\left(N^{(2)}(T)\right) + \left(\mathrm{E}(Y^{(2)})\right)^2 \mathrm{Var}\left(N^{(2)}(T)\right), \\[10pt]
    \mathrm{Cov}&\left(S_1\left(T\right), S_2\left(T\right)\right) = \mathrm{E} \left[ \left(N^{(2)}(T)\right)^2 \right] \mathrm{Cov}\left(X^{(2)}, Y^{(2)}\right)\\
    & + \mathrm{E}X^{(2)} \mathrm{Var}\left(N^{(2)}(T)\right) \mathrm{E}Y^{(2)}.
\end{align*}

\section{Wang transform}
The Wang transform is a tool from actuarial science and risk management, which helps to adjust probabilities so that they reflect risk aversion rather than just expected values \cite{Wang2002}.
Let $X$ be the future value of a financial asset at time $T$, with a cumulative distribution function $F(x)$. Wang proposed a pricing method based on a transform given as
\begin{equation}
    F^{*}(x) = \Phi \left[\Phi ^{-1}\left(F(x)\right)+\lambda \right],
    \label{eq:Wang F}
\end{equation}
where $\Phi(x)$ is a cumulative distribution function (cdf) of standard normal distribution and parameter $\lambda$ is called the market price of risk \cite{Wang1995,Wang1996a}. The mean value $E^{*}\left[X\right]$, taken under $F^{*}(x)$, defines a risk-adjusted value of $X$ at time $T$.

When applied to the cumulative distribution function of normal or log-normal distributions, Wang's transform only changes their parameters \cite{Wang2002}.
\begin{enumerate}
    \item If $F(x)$ is a cdf of normal distribution with parameters $\mu\in R$ and $\sigma>0$, denoted by $\mathcal{N}\left(\mu,\sigma^2\right)$, then $F^{*}(x)$ is a cdf of normal distribution $\mathcal{N}\left(\mu-\lambda\sigma,\sigma^2\right)$.
    \item If $F(x)$ is a cdf of log-normal distribution with parameters $\mu\in R$ and $\sigma>0$, denoted by $\mathcal{LN}\left(\mu, \sigma^2\right)$, then $F^{*}(x)$ is a cdf of log-normal distribution $\mathcal{LN}\left(\mu-\lambda\sigma,\sigma^2\right)$.
\end{enumerate}

For a liability described by the loss variable $Y$ with cdf $F(y)$, the Wang transform is given as:
\begin{equation}
    S^{*}(x) = \Phi \left[\Phi ^{-1}\left(S(x)\right)+\lambda \right],
    \label{eq:Wang S}
\end{equation}
where $S(x) = 1-F(x)$ is the tail of the distribution.

A liability can be treated as a negative asset, so two transformations can be used equivalently when dealing with insurance losses:
\begin{enumerate}
    \item apply transform given by equation \eqref{eq:Wang F} with $-\lambda$ to the cdf $F(y)$ of the loss variable;
    \item apply transform given by equation \eqref{eq:Wang S} with $\lambda$ to the cdf $F(y)$
 of the loss variable.
 \end{enumerate}

While pricing a contingent payoff $Y=h(X)$, we can apply the Wang transform in one of two ways. First, we can apply the Wang transform to the distribution $F(x)$ of the underlying risk $X$ and then find the distribution of the contingent payoff $Y^{*} = h(X^{*}),$ where $X^{*}$ is the modified underlying risk variable, after the Wang transform was applied to the distribution of $X$. Another possibility is to find the distribution $G(y)$ of the contingent payoff  for $Y=h(X)$ and apply the Wang transform to it, using the same $\lambda$ as in the first method.

In case of our models, we apply the Wang transform directly to the distributions of losses. Then, we use the modified distributions for simulations leading
to the ZC CAT bonds prices. Since we are dealing with losses, we use the transform given by formula \eqref{eq:Wang F} with negative values of the parameter $\lambda$ and investigate the impact of the parameter $\lambda$ on the final price of the ZC CAT bond.

\section{Case study: PCS data}

\begin{figure}[t]
    \centering
    \includegraphics[width=\textwidth]{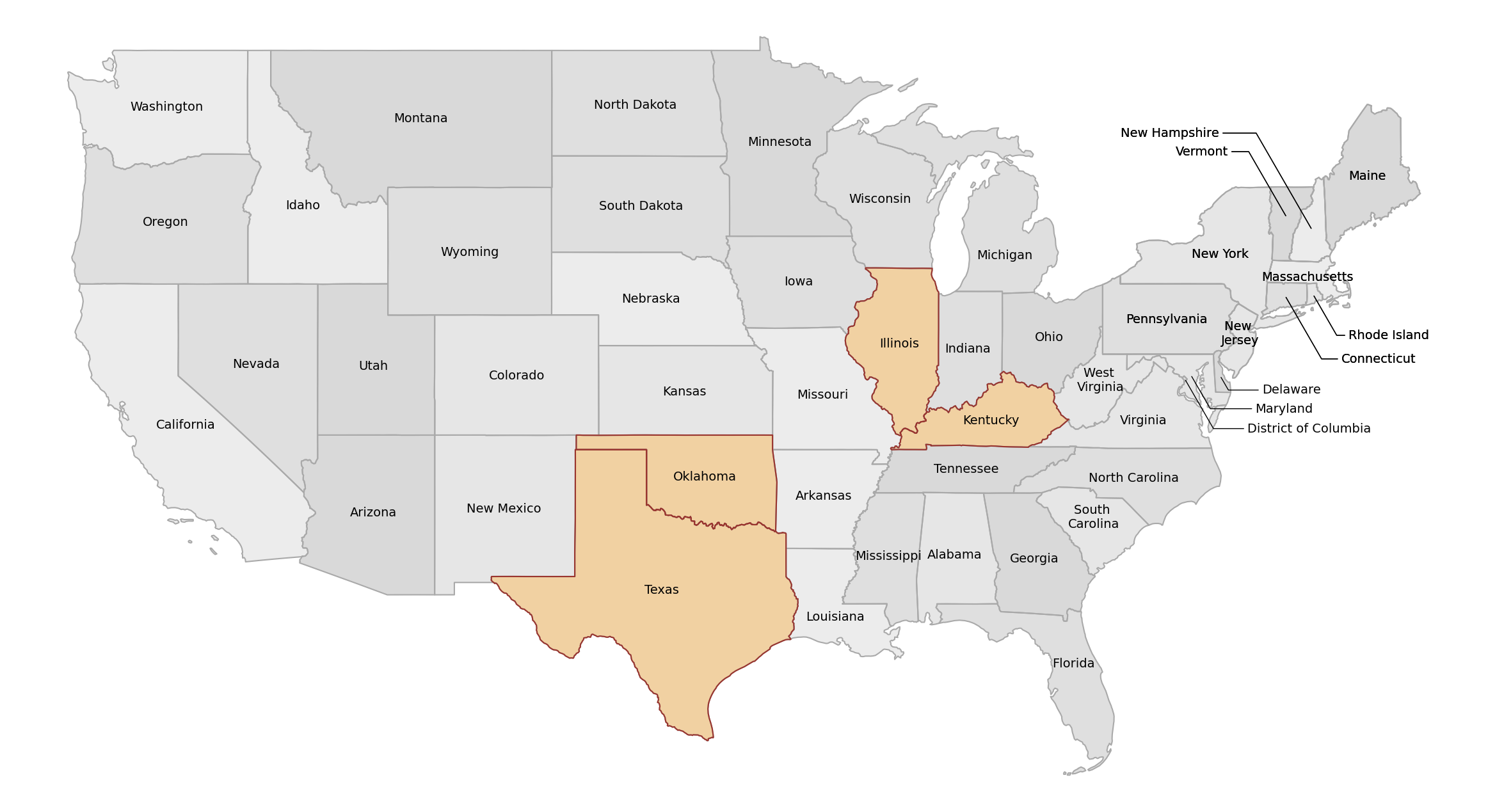}
    \caption{Location of analysed pairs of states in the USA -- Oklahoma and Texas, Illinois and Kentucky (from: \url{https://github.com/joncutrer/geopandas-tutorial}).}
    \label{fig:pairs location}
\end{figure}

Property Claim Services (PCS) is the insurance industry's well-known source for reporting catastrophic property losses. From the PCS dataset, we selected losses caused by two types of perils: wind and thunderstorm events, and winter storms, which were two of the most common perils in the set. We analyse two pairs of states: the first one is Oklahoma and Texas, the second one is Illinois and Kentucky, see Figure \ref{fig:pairs location}.

Since only natural disasters exceeding 25 million USD have recently been recorded in the PSC database, we extracted the losses from the historical database that are above that threshold (after the adjustment with the Consumer Price Index). To identify and validate the left-truncated distributions underlying the observed loss amounts, we apply the procedure explained in detail in \cite{giuricich}.

For Model 1 with independent losses, we treat the losses from each state as independent and fit their distributions separately.
Then, the sample is divided into three categories: losses that only happened in the first state, losses that only happened in the second state, and common losses caused by the same catastrophe in both states at the same time. 
In Model 2, we assume that the common losses are proportionally divided between the two states. The distribution is fitted to the total loss from two states and the $p$ of the loss is assigned to the first state and $1-p$ to the second. 
In Model 3, we fit the distributions separately to the losses from each state and calculate the Spearman correlation coefficient. We use that correlation coefficient, since, for the simulation procedure, we use the method proposed by Fackler in \cite{fackler1991genspec}, which relies on the Spearman correlation. 

\subsection{Oklahoma and Texas}
\begin{figure}[t!]
    \centering
    \includegraphics[width=\textwidth]{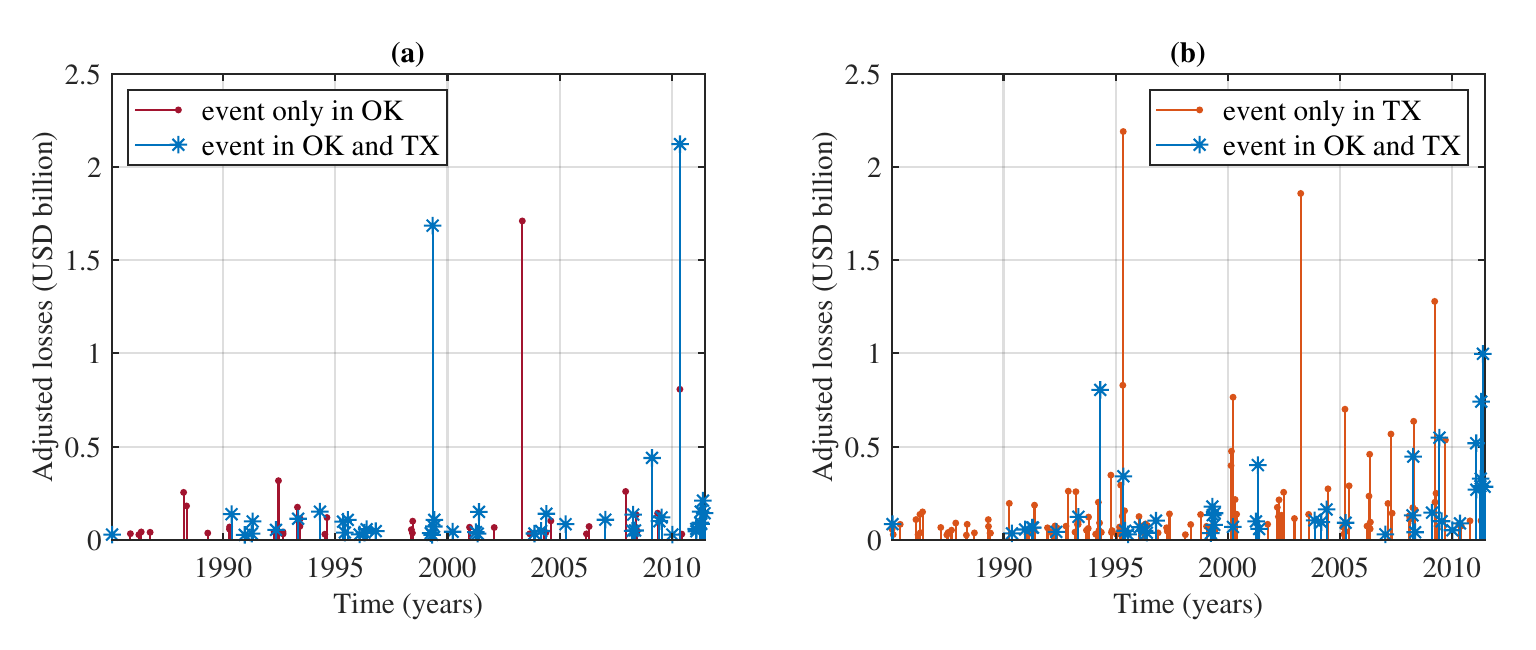}
    \caption{Adjusted losses in billion USD caused  in (a) Oklahoma, (b) Texas by wind and thunderstorm events and winter storms.}\label{fig:ok_tx_losses} 
\end{figure}

Texas and Oklahoma were the two states most affected by natural disasters (in terms of the number of the catastrophes) during the considered period. There were 163 catastrophes in Texas and 85 in Oklahoma in total, of which 44 events occurred in both states at the same time. The data set is presented in Figure \ref{fig:ok_tx_losses}.

To describe the moments of catastrophe occurrence, we fitted a homogeneous Poisson process with intensity $\lambda >0$. For the first model, we need two processes, one for all losses in Oklahoma and one for all losses in Texas. For Models 2 and  3, we need two Poisson processes counting losses that occur only in one of the states and the third one that counts the events happening in both regions. The fitted $\lambda$ values were the following:
\begin{enumerate}
    \item for all catastrophes that occurred in Oklahoma: $\lambda=2.89$;
    \item for all catastrophes that occurred in Texas: $\lambda=6.04$;
    \item for catastrophes that occurred only in Texas: $\lambda=4.76$;
    \item for catastrophes that occurred only in Oklahoma: $\lambda=1.53$;
    \item for catastrophes that hit both Oklahoma and Texas: $\lambda=1.40$.
\end{enumerate}


The next step is to identify the distribution of the losses in each case considered for the three proposed models: all losses in the region, losses that only affect the given region, common and total losses in both regions. Distributions such as log-normal, Weibull, Pareto, and inverse Gaussian were fitted to the samples, taking into account that the data were left-truncated at 25 billion dollars. The goodness of fit was tested using Kolmogorov-Smirnov (KS), Kuiper (V), Anderson Darling (AD), and Cramer von Mises (CvM) test statistics. The best results were obtained with log-normal and inverse Gaussian distributions. The fitted distributions with their parameters are presented in Table \ref{tab: OKTX dist}.

\begin{table}[t]
\caption{Results of the identification and validation procedure for losses in Oklahoma and Texas for all considered models.  Test statistics with simulated $p$-values (in italics).}
\begin{tabular}{llrrrr}
Losses & Distribution  & KS & V & AnD & CvM \\ \noalign{\smallskip}\hline\hline\noalign{\smallskip}
\multirow{2}{*}{all in OK} & Log-normal  &  0.207 & 0.353 & 0.629 & 0.074 \\
 & $\mu=-4.783, \sigma=1.841$ & \textit{0.961} & \textit{0.935} & \textit{0.642} & \textit{0.971} \\  \noalign{\smallskip}\hline\noalign{\smallskip}
 \multirow{2}{*}{all in TX} & Log-normal   & 0.523 & 0.941 & 0.499 & 0.091 \\
 & $\mu=-2.702, \sigma=1.246$ & \textit{0.862} & \textit{0.671} & \textit{0.831} & \textit{0.783} \\ \noalign{\smallskip}\hline\noalign{\smallskip}
 \multirow{2}{*}{only in OK} & Log-normal   & 0.148 & 0.272 & 0.464 & 0.057 \\
 &  $\mu=-5.012, \sigma=1.864$ & \textit{0.990} & \textit{0.967} & \textit{0.517} & \textit{0.984} \\ \noalign{\smallskip}\hline\noalign{\smallskip}
 \multirow{2}{*}{only in TX} & Log-normal   & 0.474 & 0.890 & 0.482 & 0.087 \\
 &  $\mu=-2.807,\sigma=1.266$ & \textit{0.900} & \textit{0.732} & \textit{0.860} & \textit{0.818} \\ \noalign{\smallskip}\hline\noalign{\smallskip}
total common  & Log-normal   & 0.874 & 1.480 & 0.957 & 0.156 \\
 in OK\&TX& $\mu=-1.477, \sigma=0.902$  & \textit{0.039} & \textit{0.035} & \textit{0.013} & \textit{0.020} \\ \noalign{\smallskip}\hline\noalign{\smallskip}
 \multirow{2}{*}{common in OK} & Log-normal  & 0.276 & 0.489 & 0.595 & 0.089 \\
 & $\mu=-4.564, \sigma=1.812$  & \textit{0.942} & \textit{0.885} & \textit{0.550} & \textit{0.965} \\ \noalign{\smallskip}\hline\noalign{\smallskip}
 \multirow{2}{*}{common in TX} & Inverse Gaussian & 0.496 & 0.913 & 0.244 & 0.039 \\
& $\mu=0.181,\lambda=0.098$ & \textit{0.868} & \textit{0.729} & \textit{0.916} & \textit{0.902} \\ \hline
\end{tabular}\label{tab: OKTX dist}
\end{table}

\begin{figure}[t]
    \centering
    \includegraphics[width=\textwidth]{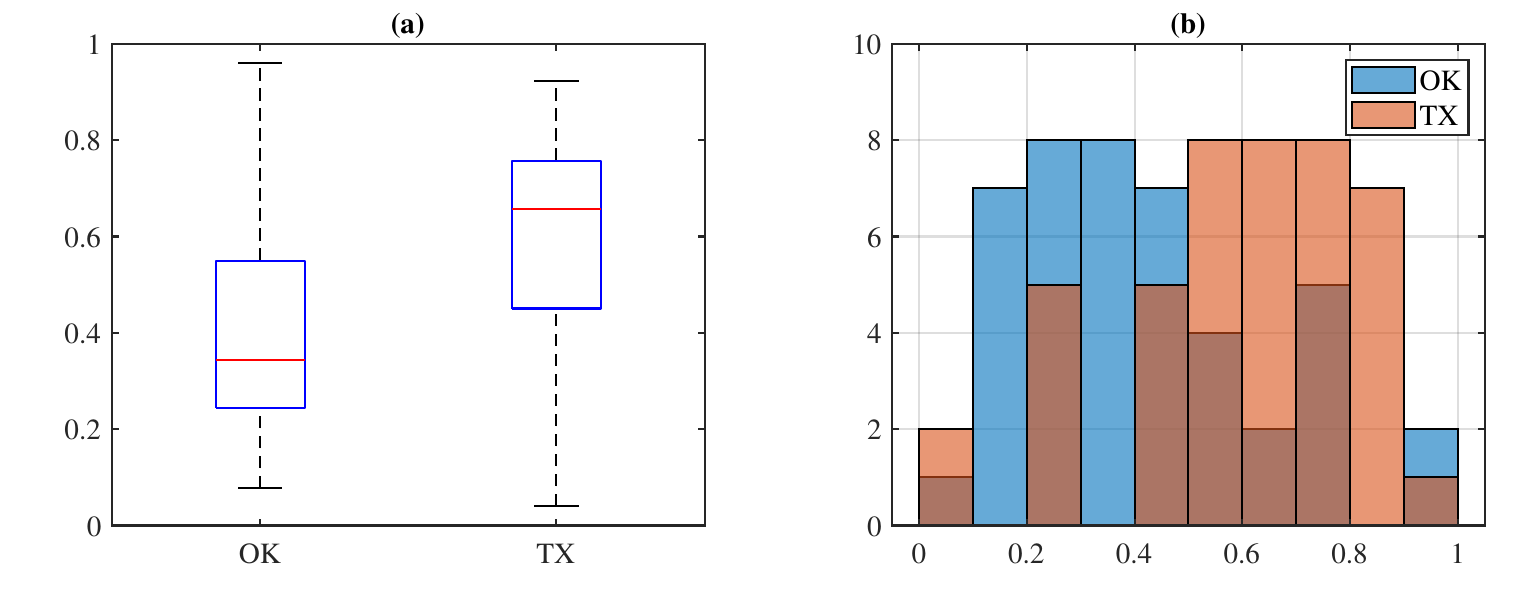}
    \caption{(a) Box plots and (b) histograms of the percentage share of common losses of Oklahoma and Texas. The average percentage share is equal to $p=0.41$.}\label{fig:ok_tx_proportion}
\end{figure}

For Models 2 and 3, we need to analyse the dependence between losses from common events. In Figure \ref{fig:ok_tx_proportion}, the percentage share of losses in Oklahoma and Texas was depicted. In most cases, the losses in Texas were higher than those in Oklahoma, although both states suffered extremely high losses a few times. Based on that analysis, the parameter $p$ of Model 2 was set to $0.41$, which corresponds to the historical mean percentage share.

In Model 3, we assume that the losses from common catastrophes are correlated with a given correlation coefficient. For the analysed data, the Spearman correlation coefficient is equal to $\rho = 0.31$, whereas the Pearson correlation coefficient is very close to zero. 

\begin{figure}[ht]
\centering\includegraphics[width=0.9\textwidth]{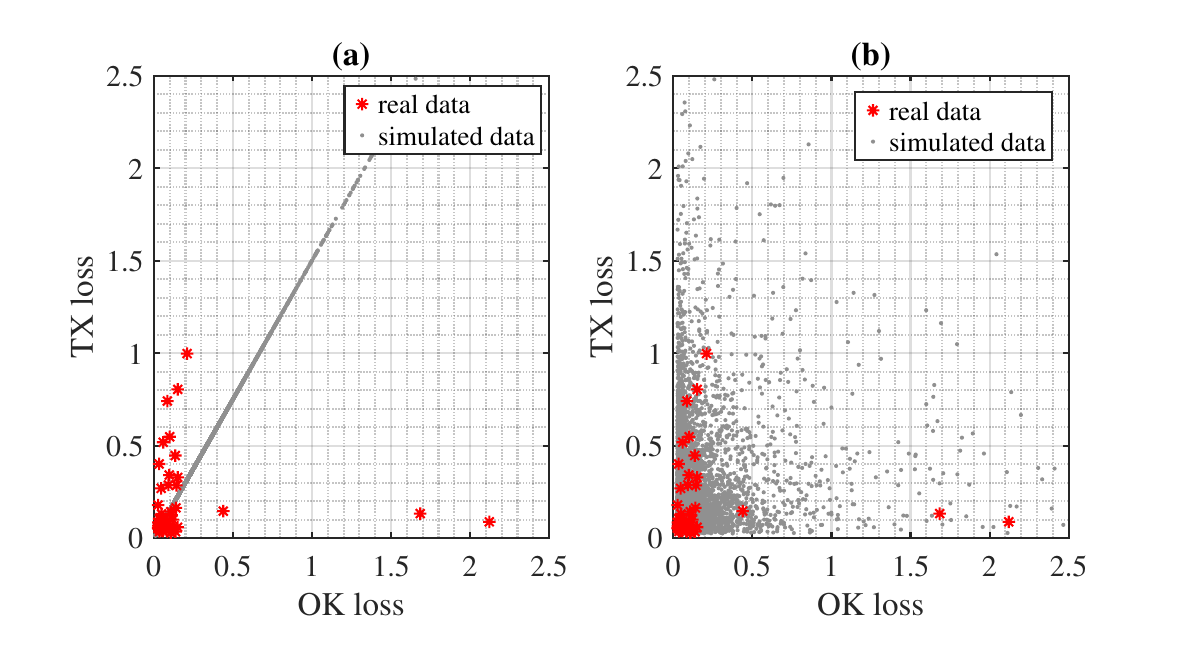}
    \caption{Comparison of real and simulated common losses in Oklahoma and Texas for (a) proportionally split losses model and (b) dependent model. }\label{fig:ok_tx_model_comparison}
\end{figure}

Both approaches to describing the losses common to both regions are illustrated in Figure \ref{fig:ok_tx_model_comparison}. The modelling approach that assumes that a total loss is split in a common proportion is shown in Figure \ref{fig:ok_tx_model_comparison}(a). We can clearly observe that it gives a worse fit than the approach that uses two correlated variables, presented in Figure \ref{fig:ok_tx_model_comparison} (b). However, assuming a constant proportion between the losses may simplify both modelling and subsequent simulations, so it is worth investigating.
 
Having fitted the models, we investigate now how the choice of the loss model and the normal approximation affect the price of a CAT bond with a payoff defined in \eqref{eq:ZC CAT payoff}. We assume the recovery rate $c=0$ and a constant interest rate $r=0.03$. The price of a two year zero-coupon CAT bond is calculated by means of 20K Monte Carlo simulations. The bond threshold for Oklahoma is denoted by $D_1$ and for Texas by $D_2$. 

\begin{figure}[hp]
\centering\includegraphics[width=\textwidth]{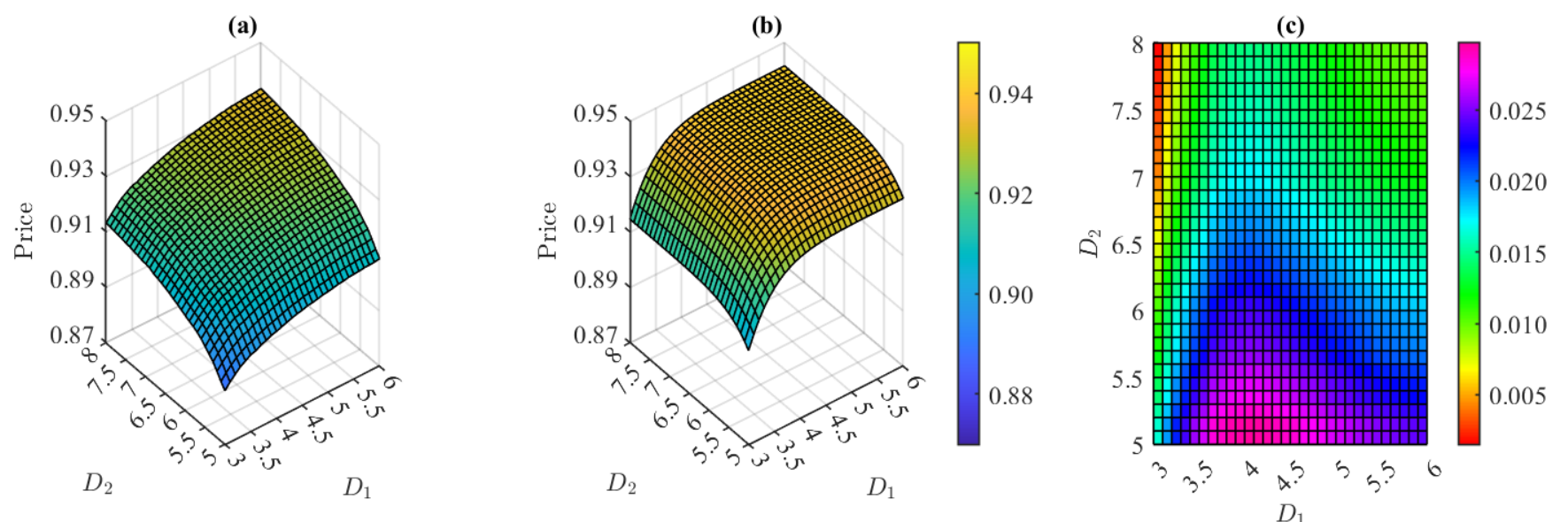}
    \caption{ZC CAT prices for independent model (Model 1) for OK and TX: (a) price from Monte Carlo simulations, (b) normal approximation and (c) relative error of the approximation.}\label{fig:ok_tx_independent_pay_app}
\centering\includegraphics[width=\textwidth]{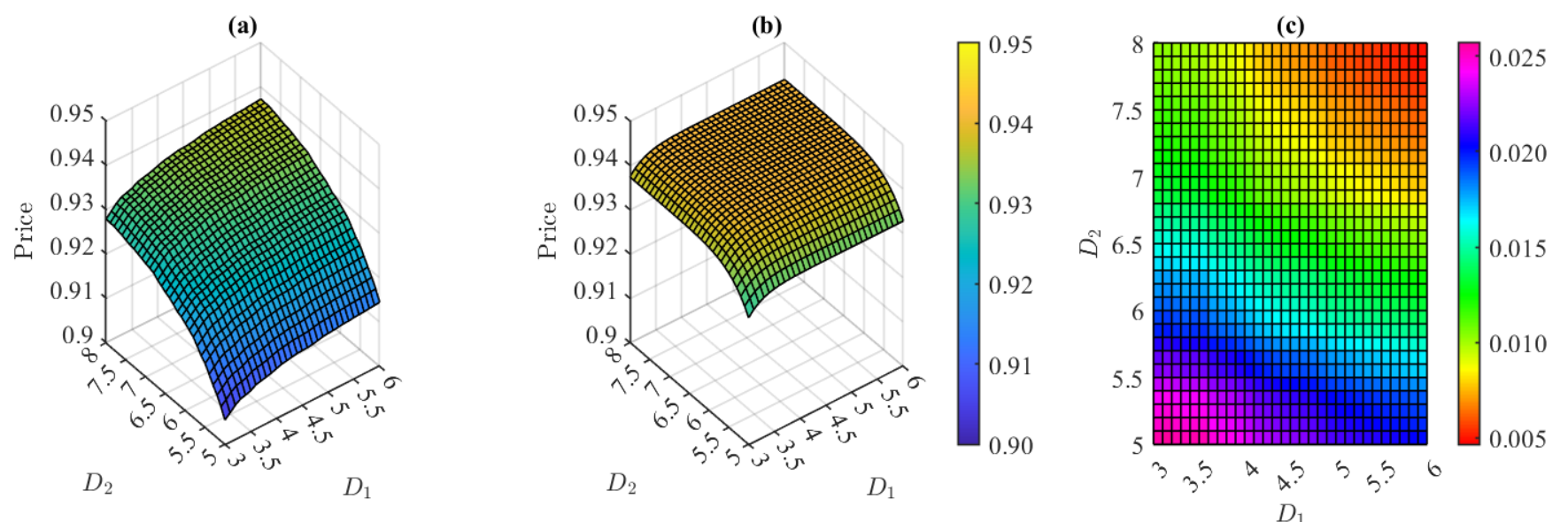}
    \caption{ZC CAT bond prices for model with proportionally split losses (Model 2) for OK and TX: (a) price from Monte Carlo simulations, (b) normal approximation and (c) relative error of the approximation.}\label{fig:ok_tx_proportional_pay_app}
\centering\includegraphics[width=\textwidth]{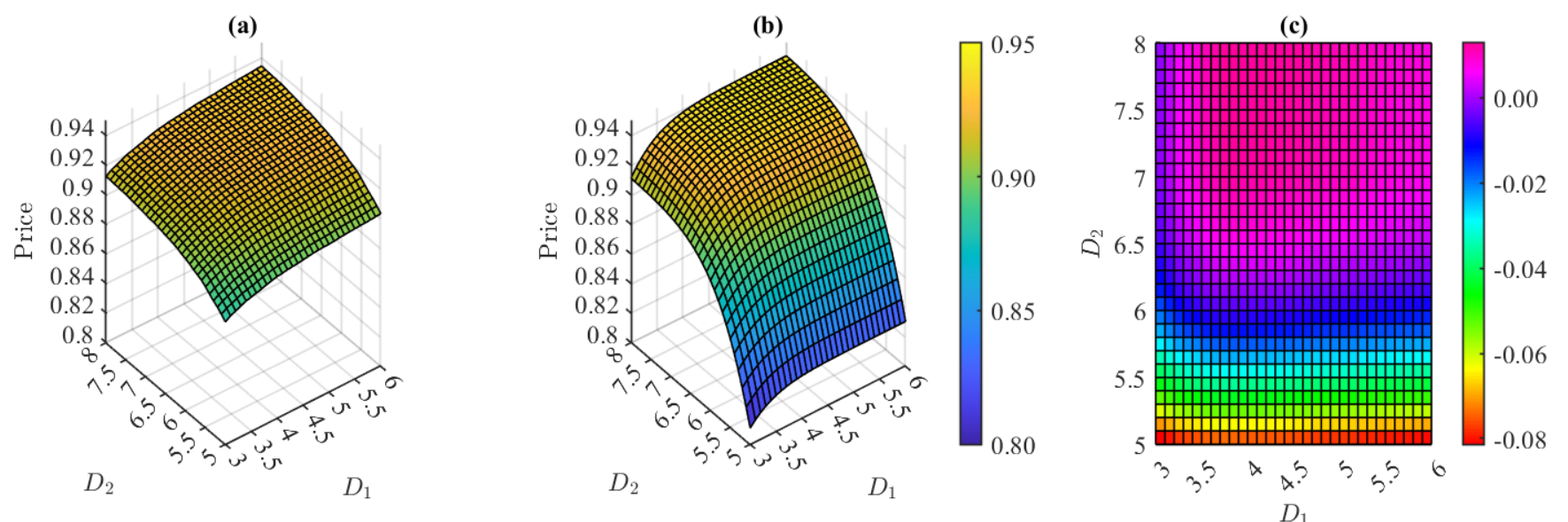}
    \caption{ZC CAT bond prices for dependent model (Model 3) for OK and TX: (a) price from Monte Carlo simulations, (b) normal approximation and (c) relative error of the approximation.}\label{fig:ok_tx_dependent_pay_app}
\end{figure}

First, the price for Model 1 with independent losses was calculated with the help of Monte Carlo simulations, see Figure \ref{fig:ok_tx_independent_pay_app}(a). As expected, the price increases with increasing thresholds $D_1$ and $D_2$. In Figure \ref{fig:ok_tx_independent_pay_app}(b) the results of the normal approximation are presented. To this end, the mean vector and covariance matrix necessary for the approximation (see equations \eqref{eq:na ind mu} and \eqref{eq:na ind sig}) were calculated for left-truncated distributions fitted to the data. We can see that the normal approximation gives rise to higher prices. The relative error of the approximation (with respect to the simulated values) is shown in Figure \ref{fig:ok_tx_independent_pay_app}(c), and does not exceed 3\%.

The relation between thresholds and the price of the CAT bond is also valid for Model 2, see Figure \ref{fig:ok_tx_proportional_pay_app}(a). The normal approximation overestimates the price (Figure \ref{fig:ok_tx_proportional_pay_app}(b)) but the relative error of the approximation again does not exceed 3\%, cf. Figure \ref{fig:ok_tx_proportional_pay_app}(c).

The normal approximation yields the worst result in the case of Model 3 with dependent losses, namely we observe larger differences between prices obtained from Monte Carlo simulations (Figure \ref{fig:ok_tx_dependent_pay_app}(a)) and the approximated values (Figure \ref{fig:ok_tx_dependent_pay_app}(b)). The approximation underestimates the price by up to 10\% in the worst case, see Figure \ref{fig:ok_tx_dependent_pay_app}(c).

\begin{figure}[t]    
\centering\includegraphics[width=\textwidth]{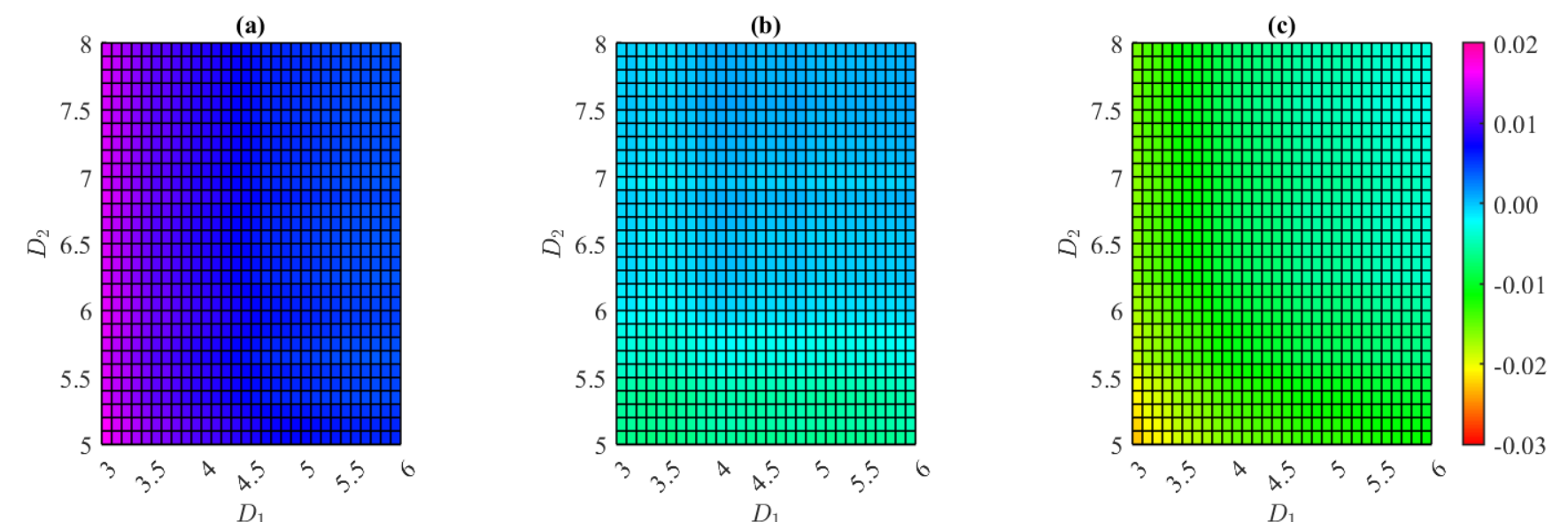}
    \caption{Differences between ZC CAT bond prices obtained from different models  for OK and TX: (a) Model 2 -- Model 1, (b) Model 3 -- Model 1, (c) Model 3 -- Model 2.}\label{fig:ok_tx_payoff_difference}
\end{figure}

Finally, the differences between the prices obtained from all models are illustrated in Figure  \ref{fig:ok_tx_payoff_difference}. We can see that Model 2 leads to higher prices than Model 1, see Figure  \ref{fig:ok_tx_payoff_difference}(a). Moreover, Model 2 gives higher prices than Model 3, see Figure  \ref{fig:ok_tx_payoff_difference}(c). Hence, Model 2 leads to the highest CAT bond prices, In Figure \ref{fig:ok_tx_payoff_difference}(b) we can observe that the differences between the prices obtained from Model 1 and Model 3 are mostly close to zero, only for lower values of $D_2$ the prices calculated from Model 1 are significantly higher.

\subsection{Illinois and Kentucky}

\begin{figure}[t]
    \centering
    \includegraphics[width=\textwidth]{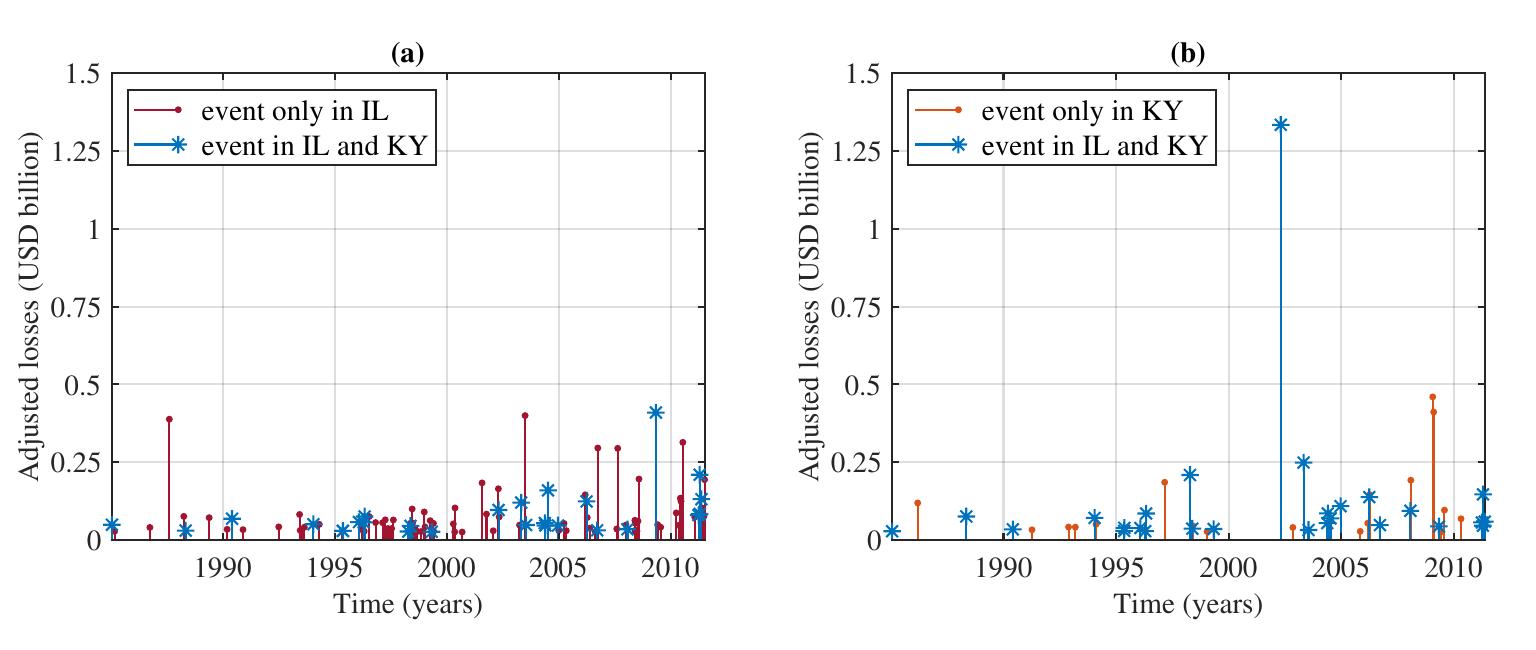}
    \caption{Adjusted losses in billion USD caused in (a) Illinois, (b) Kentucky by wind and thunderstorm events and winter storms.}\label{fig:il_ky_losses}    
    \centering
    \includegraphics[width=\textwidth]{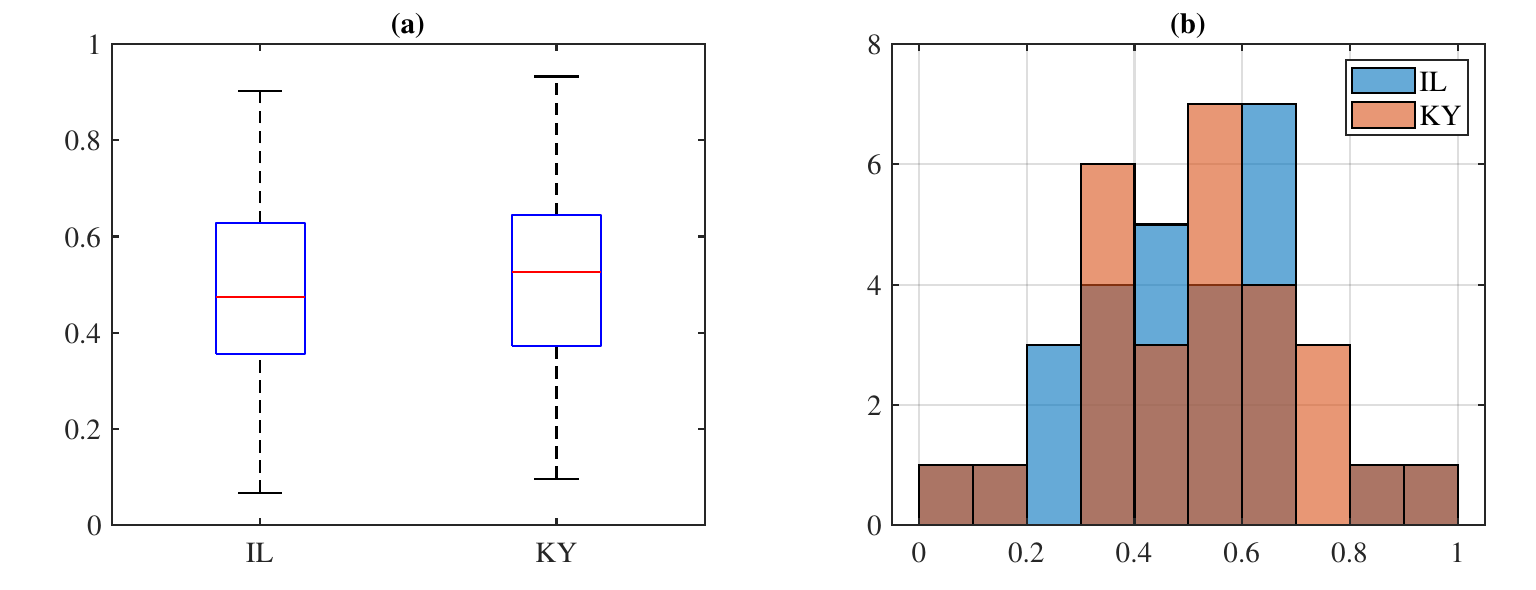}
    \caption{(a) Box plots and (b) histograms of the percentage share of common losses of Illinois and Kentucky. The average percentage share is to $p=0.49$ for Oklahoma.}\label{fig:il_ky_proportion}
\end{figure}

We now study the natural catastrophe losses that occurred in Illinois and Kentucky. In Illinois 111 catastrophes occurred, and Kentucky was affected by 45 events in the analysed period. Of these, 18 events affected both states simultaneously.  The adjusted losses in both states are presented in Figure \ref{fig:il_ky_losses}.

Similarly as before, five counting processes are needed, two for all losses in each region, two for losses affecting only one region, and one process for losses from disasters occurring in both regions. We fitted intensities of homogeneous Poisson processes and obtained the following results:
\begin{enumerate}
    \item for all catastrophes that occurred in Illinois: $\lambda=3.59$;
    \item for all catastrophes that occurred in Kentucky: $\lambda=1.46$;
    \item for catastrophes that occurred only in Illinois: $\lambda=2.72$;
    \item for catastrophes that occurred only in Kentucky: $\lambda=0.59$;
    \item for catastrophes that hit both Illinois and Kentucky: $\lambda=0.89$.
\end{enumerate}

The same left-truncated distributions as for losses in Oklahoma and Texas were tested for Illinois and Kentucky. The distribution parameters that gave the best fit, together with test statistics and simulated $p$-values, are presented in Table \ref{tab: ILKY dist}. The distributions of common losses in Illinois and Kentucky were best described by the Pareto and log-normal distributions, respectively. The log-normal distribution was selected for all other cases.

\begin{table}[t]
\caption{Results of the identification and validation procedure for losses in Illinois and Kentucky for all considered models.  Test statistics with simulated $p$-values (in italics).}
\begin{tabular}{llrrrr}
Losses & Distribution& KS & V & AnD & CvM \\ \noalign{\smallskip}\hline\hline\noalign{\smallskip}
\multirow{2}{*}{all in IL} & Log-normal  & 0.215 & 0.390 & 0.544 & 0.075 \\
& $\mu=-4.554, \sigma=1.386$ & \textit{0.962} & \textit{0.915} & \textit{0.667} & \textit{0.976} \\ \noalign{\smallskip}\hline\noalign{\smallskip}
\multirow{2}{*}{all in KY} &Log-normal & 0.070 & 0.1370 & 0.119 & 0.014 \\
 &  $\mu=-4.869, \sigma=1.736$  & \textit{1.000} & \textit{1.000} & \textit{0.526} & \textit{1.000} \\ \noalign{\smallskip}\hline\noalign{\smallskip}
 \multirow{2}{*}{only in IL} & Log-normal   & 0.118 & 0.197 & 0.468 & 0.063 \\
 & $\mu=-5.288. \sigma=1.569$ & \textit{0.982} & \textit{0.973} & \textit{0.555} & \textit{0.989} \\ \noalign{\smallskip}\hline\noalign{\smallskip}
 \multirow{2}{*}{only in KY} & Log-normal   & 0.120 & 0.219 & 0.271 & 0.037 \\
 & $\mu=-4.709,\sigma=1.749$  & \textit{0.998} & \textit{0.994} & \textit{0.506} & \textit{0.994} \\ \noalign{\smallskip}\hline\noalign{\smallskip}
total common & Log-normal   & 0.641 & 1.150 & 0.621 & 0.086\\
 in IL\&KY&  $\mu=-1.942, \sigma=0.718$ & \textit{0.327} & \textit{0.252} & \textit{0.111} & \textit{0.121} \\ \noalign{\smallskip}\hline\noalign{\smallskip}
 \multirow{2}{*}{common in IL} & Pareto  & 0.307 & 0.518 & 0.228 & 0.037 \\
 &  $k=0.314,\sigma=0.032$ & \textit{0.918} & \textit{0.867} & \textit{0.86} & \textit{0.98} \\ \noalign{\smallskip}\hline\noalign{\smallskip}
 \multirow{2}{*}{common in KY} & Log-normal  & 0.106 & 0.187 & 0.283 & 0.029 \\
 & $\mu=-4.918,\sigma=1.713$ & \textit{0.998} & \textit{0.995} & \textit{0.499} & \textit{0.998} \\ \noalign{\smallskip}\hline\noalign{\smallskip}
\end{tabular}\label{tab: ILKY dist}
\end{table}

\begin{figure}[t]   \centering\includegraphics[width=0.8\textwidth]{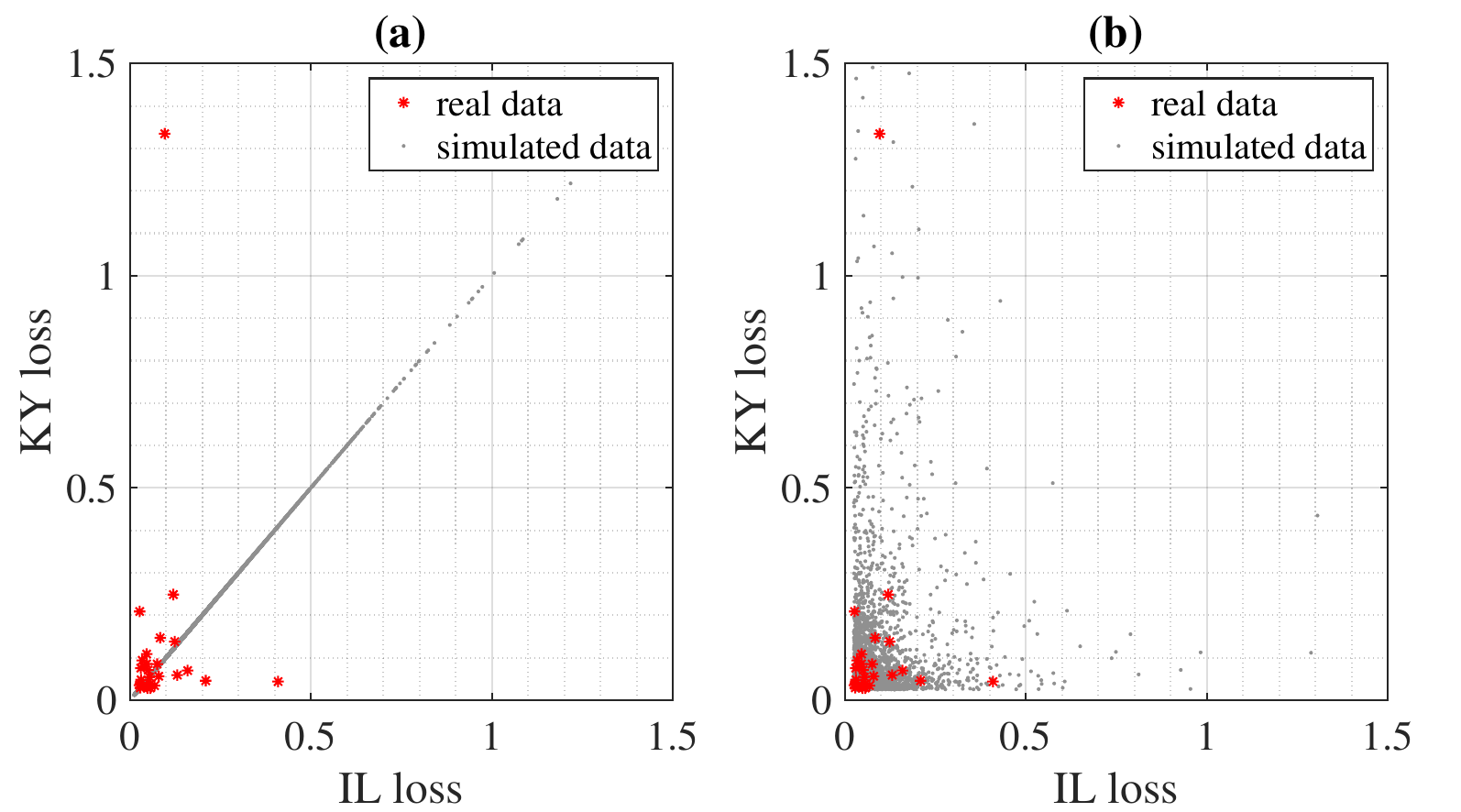}
    \caption{Comparison of real and simulated common losses in Illinois and Kentucky for (a) proportionally split losses model and (b) dependent model. }\label{fig:ilky_common_losses_model_comparison}
\end{figure}

The distribution of the percentage share of common losses of
Illinois and Kentucky is depicted in Figure \ref{fig:il_ky_proportion}. Compared to the previous pair, in this case, the losses are more evenly distributed across two states. The average percentage share was close to $p=0.5$, which is the value chosen for Model 2. The Spearman correlation coefficient is equal to $\rho = 0.22$, whereas the Pearson correlation coefficient is very close to zero.
Two approaches to modelling losses from common events in Illinois and Kentucky are illustrated in Figure \ref{fig:ilky_common_losses_model_comparison}, with the approach of Model 3 giving a better fit to the data than the approach of Model 2.

In order to obtain the CAT bond prices, we choose the same bond parameters as before, so $c=0$ and $r=0.03$. The price of a two year zero-coupon CAT bond is calculated by means of 20K Monte Carlo simulations. The bond threshold for Illinois is denoted by $D_1$ and for Kentucky by $D_2$.

\begin{figure}[ht]    \centering\includegraphics[width=\textwidth]{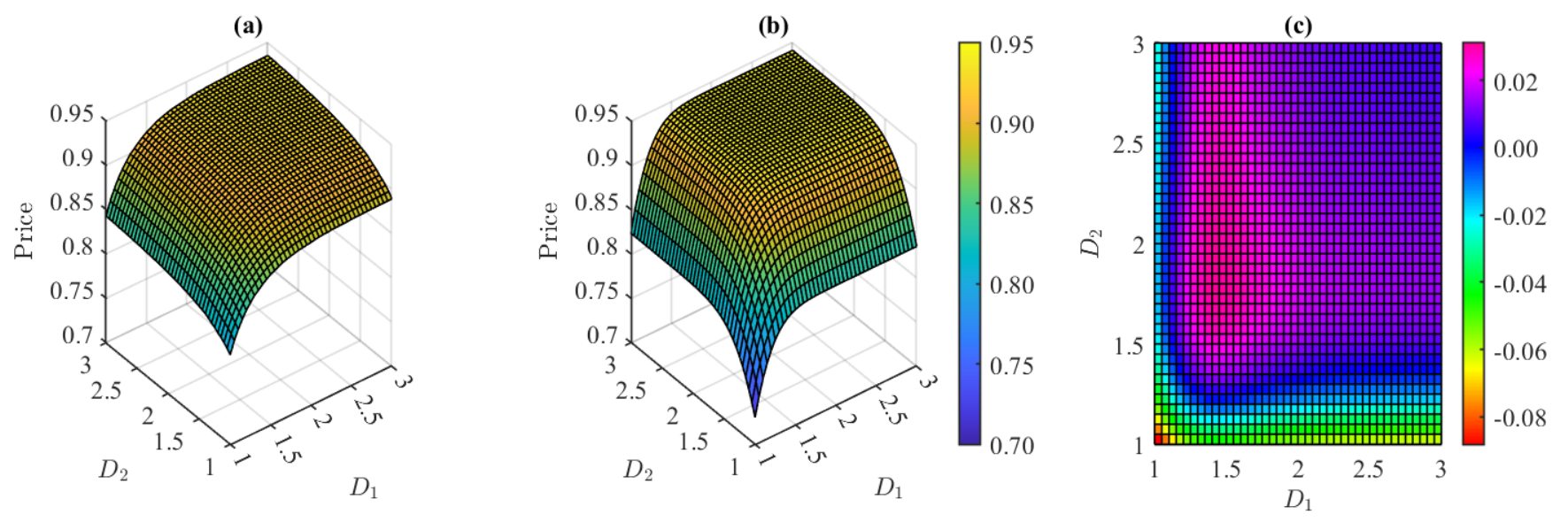}
    \caption{ZC CAT bond prices for independent model (Model 1) for IL and KY: (a) price from Monte Carlo simulations, (b) normal approximation and (c) relative difference.}\label{fig:ilky_independent_approx}
\centering\includegraphics[width=\textwidth]{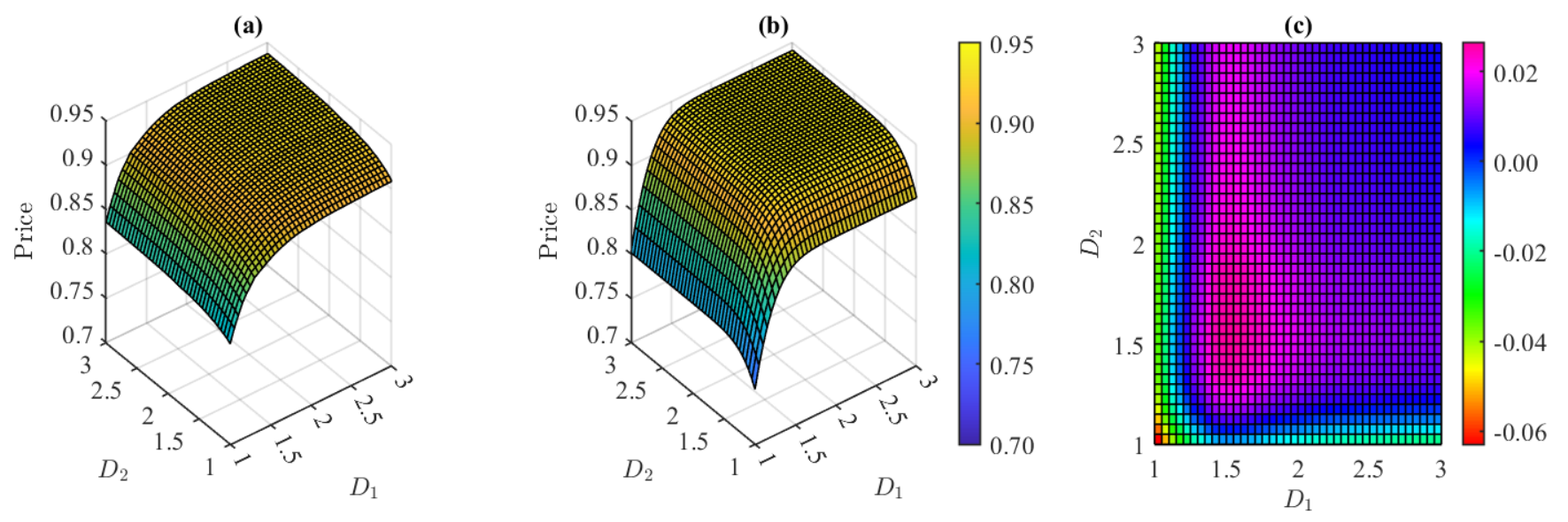}
    \caption{ZC CAT bond prices for model with proportionally split losses (Model 2) for IL and KY: (a) price from Monte Carlo simulations, (b) normal approximation and (c) relative difference.}\label{fig:ilky_proportional_approx}
    \end{figure}

The prices for Model 1 with independent losses calculated with the help of Monte Carlo simulations are presented in Figure \ref{fig:ilky_independent_approx}(a). The price clearly increases with increasing thresholds $D_1$ and $D_2$. In Figure \ref{fig:ilky_independent_approx}(b) the results of the normal approximation are presented. We can see that, contrary to the previously considered pair of states, the normal approximation mostly yields lower prices. The relative error of the approximation (with respect to the simulated values) is presented in Figure \ref{fig:ilky_independent_approx}(c), and reaches -8\%.

The prices calculated for Model 2 are shown in Figure \ref{fig:ilky_proportional_approx}(a). They clearly increase with respect to the thresholds $D_1$ and $D_2$. Now, the normal approximation is quite close to the price obtained from Monte Carlo simulations, see Figures \ref{fig:ilky_proportional_approx}(b) \ref{fig:ilky_proportional_approx}(c). The relative error of the approximation again usually is low only for small $D_1$'s and $D_2$'s it reaches -6\%.

\begin{figure}[ht] 
\centering\includegraphics[width=\textwidth]{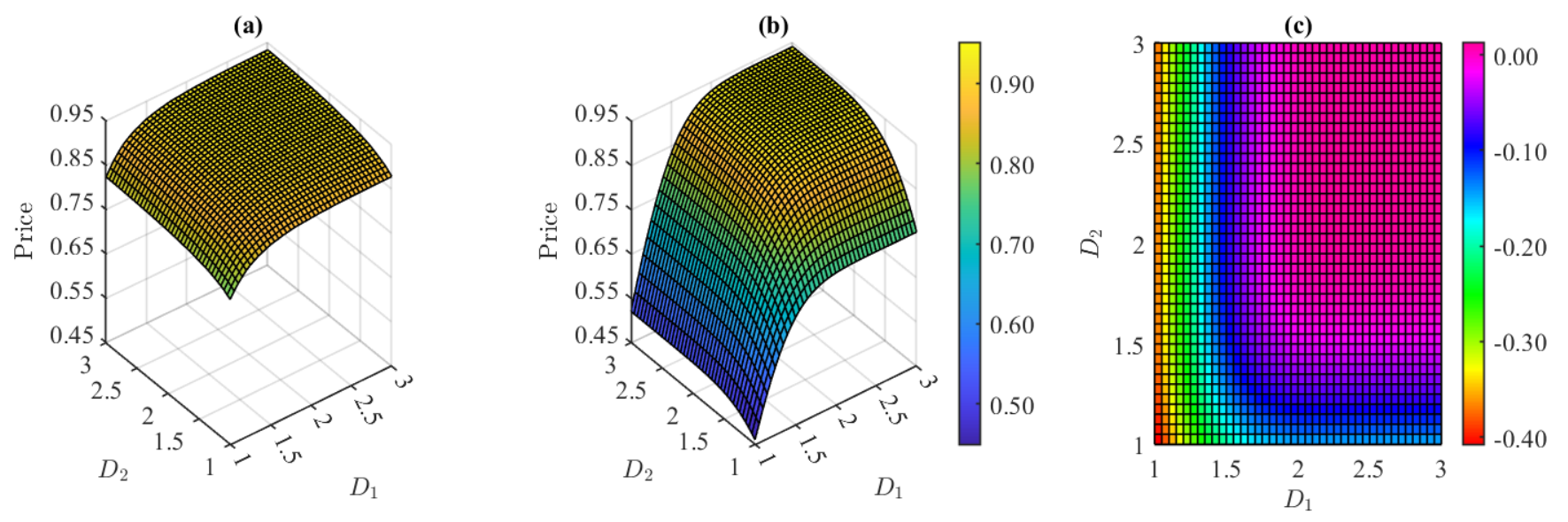}
    \caption{ZC CAT bond prices for dependent model (Model 3) for IL and KY: (a) price from Monte Carlo simulations, (b) normal approximation and (c) relative difference.}\label{fig:ilky_dependent_approx}    
\end{figure}

As for the previous pair of states, the normal approximation yields the worst result in the case of Model 3 with dependent losses, namely we observe larger differences between prices obtained from Monte Carlo simulations (Figure \ref{fig:ilky_dependent_approx}(a)) and the approximated values (Figure \ref{fig:ilky_dependent_approx}(b)). The approximation underestimates the price by up to 40\% in the worst case, cf. Figure \ref{fig:ilky_dependent_approx}(c).



\begin{figure}[ht!]
\centering\includegraphics[width=\textwidth]{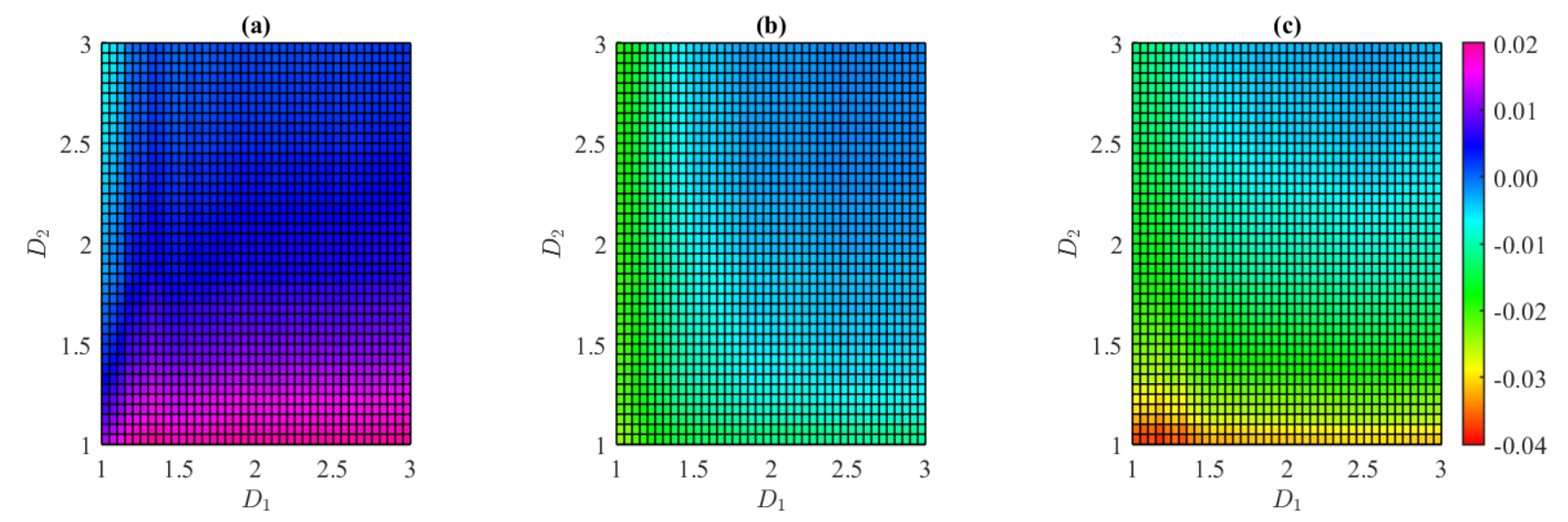}
    \caption{Differences between ZC CAT bond prices obtained from different models for IL and KY: (a) Model 2 -- Model 1, (b) Model 3 -- Model 1, (c) Model 3 -- Model 2.}\label{fig:il_ky_payoff_difference}
\end{figure}

To compare all three models, in Figure \ref{fig:il_ky_payoff_difference} we analyse the differences between the obtained prices. From Figure \ref{fig:il_ky_payoff_difference}(a), we can see that Model 2, including the relation between common losses, usually leads to higher prices. We can observe in Figure \ref{fig:il_ky_payoff_difference}(b) that Model 3 gives lower prices than Model 1. We can see in Figure \ref{fig:il_ky_payoff_difference}(c)  that all differences are negative, which means that the prices obtained from Model 2 are also higher than the prices obtained from Model 3, so, in this case, including the correlation slightly decreases the price of the bond.

\subsection{Wang transform}
\begin{figure}[ht]    
\centering\includegraphics[width=\textwidth]{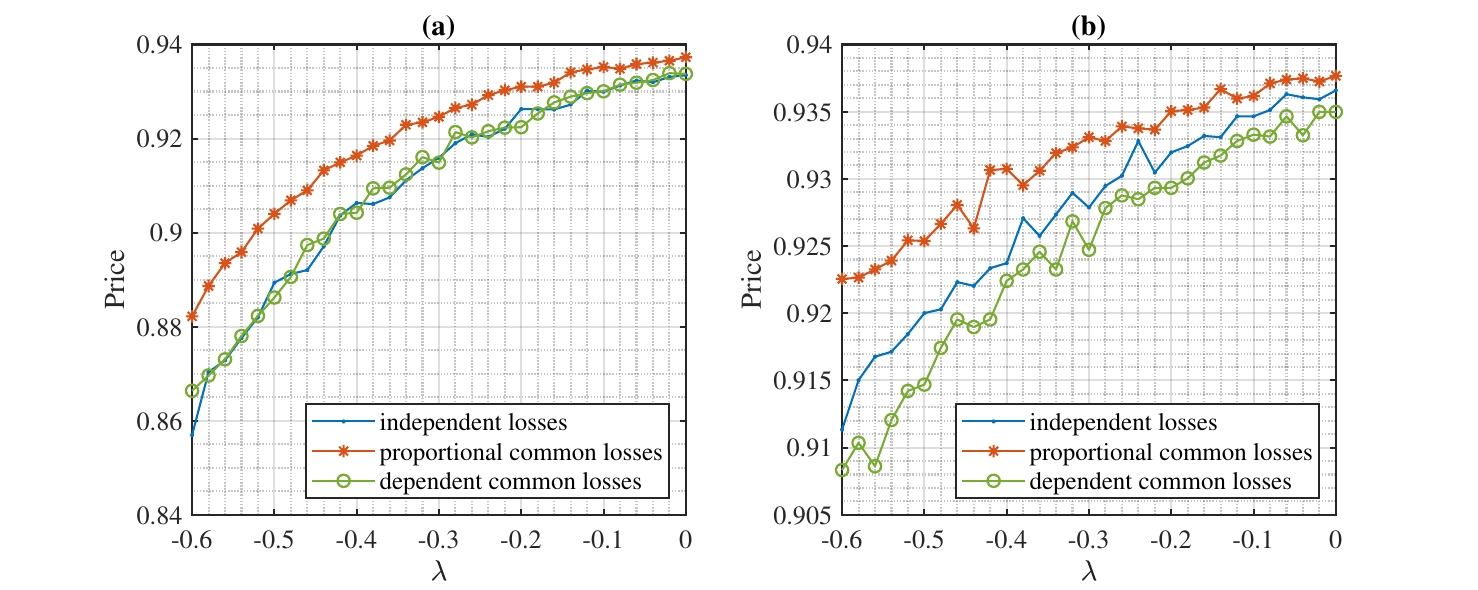}
    \caption{Prices of ZC CAT bond obtained from different models (by means of Monte Carlo simulations), for (a) OK and TX, (b) IL and KY, with Wang transform applied to the distribution of losses with varying parameter $\lambda.$}\label{fig:wangtransfrom}

\end{figure}
In this part, we examine the influence of the Wang transform on the CAT bond pricing under each model. We applied the transform defined in \eqref{eq:Wang F} directly to the distribution of the losses, so negative values of the parameter $\lambda$ are used. 

In the case of Oklahoma and Texas, we set the thresholds at $D_1 = 6$ and $D_2 = 8$ billion dollars. The results for all models are shown in Figure \ref{fig:wangtransfrom}(a). With an increase in the absolute value of the parameter $\lambda$, representing the market price of risk, the price of the CAT bond decreases. We also notice that for the chosen values of $D_1$ and $D_2$, Model 2 gives the highest prices, while Model 1 and Model 3 give similar values. 

Similar results were obtained for Illinois and Kentucky, see Figure \ref{fig:wangtransfrom}(b). In this case, both thresholds were equal to 3 billion dollars. For the chosen values of $D_1$ and $D_2$, Model 2 gives the highest prices, and the prices obtained from Model 1 are higher than those from dependent losses.

\section{Conclusions}




Catastrophe bonds have emerged as a vital mechanism for transferring disaster risk to capital markets, offering benefits such as portfolio diversification, improved risk management for insurers, and improved financial stability for disaster-prone regions \cite{Barrieu2010}.However, the complexities in pricing, structuring, and market accessibility highlight challenges that continue to shape the evolution of the CAT bond market. As climate change amplifies the frequency and severity of catastrophic events, the role of CAT bonds is expected to grow, prompting ongoing research to improve their efficacy and expand their use in both financial and societal risk-management strategies \cite{burnecki2023catastrophe}. 

In this study, we proposed and analysed three distinct modelling approaches for representing catastrophe losses across multiple geographic regions. The models capture a spectrum of dependence structures: from full independence (Model 1), to proportional sharing of common losses (Model 2), to flexible correlation-based dependence (Model 3). 

Our results show that the assumed dependency  mechanism significantly affects the valuation of the CAT bond. In particular, Model 2 systematically yields higher prices than both Models 1 and 3, reflecting its more conservative treatment of common catastrophic events. Moreover, Model 3 can generate higher or lower prices depending on the empirical correlation structure; for the datasets studied (Oklahoma–Texas and Illinois–Kentucky), introducing correlation tended to decrease prices relative to the proportional model, although its performance varied between thresholds. 

We further evaluated a normal approximation to the two-dimensional loss distribution. Across all models, the approximation was computationally efficient and reasonably accurate for moderate threshold values, typically producing errors below 3\% for the first two models. However, for the more complex dependent-loss structure (Model 3), the approximation deteriorated, especially for the Illinois–Kentucky dataset, where errors reached 40\%. This highlights that reliance on normal approximations should be applied with caution when modelling correlated heavy-tailed phenomena, especially in the presence of left-truncation and heterogeneous dependence. 


We also analysed the influence of the Wang transform to incorporate a market price of risk. In both regional case studies, increasing the absolute value of the distortion parameter led to lower CAT bond prices, consistent with the transform’s risk-adjustment effect. The ranking of models remained stable: the proportional-loss model consistently produced the highest prices, whereas the dependent-loss model yielded prices closest to the independent case. These findings confirm that the choice of dependence assumptions plays a more dominant role than risk-loading adjustments in shaping final price levels. 

Overall, our framework demonstrates how multi-regional loss modelling can substantially influence CAT bond valuation, and it provides practitioners with flexible tools to incorporate different regional interdependencies. While the study focuses on two-region settings, the modelling concepts naturally extend to higher-dimensional problems, including multi-peril or multi-state ILS structures. 


We believe that our findings can be helpful in modelling and pricing other ILS tied to natural disasters and provide a foundation for further research toward more accurate and robust multi-region catastrophe-risk valuation. 

\section*{Acknowledgements}
The authors’ research was supported by the National Science Centre, Poland (NCN), OPUS grant no. 2022/47/B/HS4/02139.

\bibliographystyle{elsarticle-harv} 
\bibliography{bibliography}

\end{document}